\documentclass[preprint,showpacs,preprintnumbers,amsmath,amssymb]{revtex4}

\usepackage{graphicx}% Include figure files
\usepackage{dcolumn}% Align table columns on decimal point
\usepackage{bm}% bold math

\begin{document}

%\preprint{APS/123-QED}

\title{Superconducting properties of the 2D models\\
       with different types of inter-particle coupling}

\author{V.M.~Loktev} \affiliation{Bogolyubov Institute for Theoretical Physics,
 Metrologicheskaya Str. 14-b, Kiev, 03143, Ukraine}%
 \email{vloktev@bitp.kiev.ua} 
 \author{V.~Turkowski} \email{vturk@cfif.ist.utl.pt}
%\homepage{http://www.Second.institution.edu/~Charlie.Author}
 \affiliation{CFIF, Instituto Superior Tecnico,
         Av.Rovisco Pais, 1049-001 Lisbon, Portugal}

\date{April 7, 2003}% It is always \today, today, but you may specify any
% date with \date.

\begin{abstract}
The superconducting properties of the 2D fermion system
with local and different types of the
indirect boson-exchange attractions
in the cases of $s$-wave and $d$-wave pairing
are reviewed and analysed at $T=0$. In particular, the possibility of the
crossover from the
Bose-Einstein condensation regime to the BCS-like 
superconductivity with the carrier density and coupling changing
in the case of different pairing channels is discussed.
Gaussian fluctuations of the order parameter are taken into account
and the carrier density dependence of the gap is studied in this case.
The role of the interaction form 
in the physical behavior of the systems is also discussed.
\end{abstract}


% 74.20.Rp - Pairing symmetries (other than s-wave)
% 74.25.Dw - Superconductivity phase diagram 
% 74.62.Dh - Effects of crystal defects, doping and substitution
% 74.72.-h - High-T_c compounds

\maketitle

\section{\label{sec:level1} Introduction}

The theoretical description of the carrier density dependence of
the physicsl properties of superconductors has a long history. 
Probably, the first attempt
to solve this problem self-consistently was made by Eagles in
Ref. \cite{Eagles}, where the author tried to apply his results
for description of the superconducting properties of $Zr$-doped $SrTiO_{3}$
\cite{Frederikse}. The author studied 
the dependence of the superconducting gap at $T=0$ and 
of the mean-field critical temperature on charge carrier density
in the two-dimensional (2D) and in the three-dimensional (3D) 
systems with phonon-like attraction at low carrier
densities.
The set of the equations for the superconducting gap and
Fermi energy in two and three dimensions was analyzed.
It was estimated that the diameter of pairs is smaller than
distance between pairs
at low carrier densities, therefore, superconductivity at
low carrier densities transforms into superfluidity of small (local) pairs.
Now this phenomena is known as crossover from superfluidity
to superconductivity with carrier density changing.
It was also shown by Eagles that there exists a critical value
of the inter-particle coupling below which there is no local
pairs at low carrier densities in the 3D system.
It is worth to mention that the possibility of superfluidity scenario
of superconductivity was also proposed
by Ogg \cite{Ogg} and Schafroth \cite{Schafroth1,Schafroth2,Schafroth3}
before paper \cite{Eagles}. 

Later the problem of the crossover was studied by Leggett
in \cite{Leggett}, where the author studied 3D system with short-range
repulsion and finite-range attraction. He analyzed the properties
of the system changing dimensionless parameter $1/(ak_F)$,
where $a$ is the scattering length and $k_F$ is the Fermi momentum
which defines the particle number in the system. It was shown
that in the limiting cases 
the system consists of bound ``diatomic molecules'' at $1/(ak_F)=\infty$,
and the Cooper pairing takes place at $1/(ak_F)=-\infty$.

Nozi$\grave{e}$res and Schmitt-Rink \cite{Nozieres} 
generalized the results of Leggett on the cases of finite temperatures
and lattice model when
the interaction potential is separable.
They have shown that the crossover from superfluidity to superconductivity
is smooth with $k_{F}$ and interaction (attraction) potential $V$ changing,
when $V$ is larger than the critical value for two-particle bound state
formation.
All the results mentioned above correspond to the case of the 
$s$-wave pairing.

The real boom in the interest to the crossover phenomena has started
in the second half of eighties after the discovery of high-temperature superconductors (HTSCs),
materials with very unusual dependence of superconducting properties 
on the carrier density.

Theory of the HTSC still remains
one of the most difficult and one of the most important
problems of the modern condensed matter theory.
Due to complicated crystal structure, anysotropic pairing,
strong electron correlations, presence of the disorder, low dimensionality
of this materials and some other complications, the construction
of the self-consistent theory of HTSCs is not completed
so far.

During the last years many models which take into account some of the
cuprate properties were proposed for describing
the doping and interaction dependence of the superconducting gap and
of the superconducting critical temperature.

The 3D crossover from superfluidity to supercondictivity
in the $s$-wave pairing channel for the model
with local attraction was studied in
\cite{SadeMelo1,Haussmann1,Randeria3,Micnas1,Engelbrecht1,Marini1,Andrenacci1,Babaev1} 
for the model with local (so called 4F-) attraction, and in \cite{Pistolesi2} for the case
of separable potential.
In \cite{Micnas1} the additional case with local repulsion
and inter-site attraction was considered.
The role of the order parameter fluctuations
at $T=0$ was analyzed in \cite{Aitchison1} (3D case) and 
in \cite{Kos1} (2D and 3D cases). 
The quasi-2D model with $s$-wave pairing at $T=0$
was considered in \cite{Gorbar1}.

For the 2D case the problem of the crossover in the $s$-wave pairing channel
was analyzed
at zero temperature (when a long-range    
superconducting order is still possible in a 2D system 
\cite{Mermin1}) in \cite{Randeria1,Randeria2,Micnas1,Andrenacci1,Pistolesi1,Pistolesi2,Fehrenbacher1,Randeria3,Nozieres1,Gorbar2}, 
and at finite temperatures in \cite{Micnas1,Pistolesi1,Pistolesi2,Fehrenbacher1,Randeria3,Borkowski1,Nozieres1,Andrenacci1,Duncan1,Babaev1}. 
The metal-superconductor phases boundaries on the $n-U$ phase
diagram of the 2D attractive Hubbard model at $T=0$ 
was studied in \cite{Capone1} by means of the Dynamical Mean-Field Theory.
The crossover problem for the 2D phonon-mediated model in the $s$-wave
channel at T=0 was studied in \cite{JPS,TMF}.

The 2D $d$-wave crossover for models with separable potential was studied in \cite{Borkowski1,denHertog1,Andrenacci1} at $T=0$. 
The Ginzburg-Landau potential with carrier density dependent parameters
for the 2D $d$-wave superconductor at finite $T$ was considered in 
\cite{Stintzing1}. The doping dependence of the critical temperature 
in the strongly
correlated electron model with the electron-phonon coupling was studied in
\cite{Mierzejewski1,Paci1}. It was shown that the vertex correction
to the electron-phonon interaction in this model leads
to the increasing of superconductivity in the $d$-wave pairing channel.
In paper \cite{Fehrenbacher1} the cases with different 
pairing symmetries $s,s_{xy},d_{xy}$ and $d_{x^{2}-y^{y}}$
in the 2D system with local, nearest neighbor, and
next nearest neighbor attraction were studied at $T=0$ and finite $T$
in order to describe
the doping dependence of the zero superconducting gap and superconducting 
critical temperature. The possibility of zero temperature 
$s$-, $d$- and mixed $s+id$ pairing
in the 2D system 
as function of coupling were studied in paper \cite{Musaelian1}.
The same problem in the quasi-2D Hubbard model with nearest
neighbor attraction at finite temperature was studied
in \cite{Wallington1}, where also doping dependence of the superconducting
properties was considered.

The 2D model with the nearest neighbor (n.n.) attraction and next nearest
neighbor (n.n.n.) hopping at $T=0$ was considered
in \cite{Soares1}. 
It was shown that at some relation between
n.n. and n.n.n. hopping the system is always in Bose-Einstein condensation
regime and that there is no pairing at low carrier
density when coupling is weak (see also \cite{denHertog1}).
However, as it was shown in \cite{Perali1} this statement
is not correct.
The 2D model with local repulsion and n.n. attraction at $T=0$
in the $s$-wave and $d$-wave channels was studied in \cite{Pistolesi3}.
The role of the n.n.n. hopping was analyzed.
The $s$-wave and $d$-wave pairing crossover at $T=0$
in the model with doping dependent attraction was considered in \cite{magnons}.
It is interesting to mention that the boson-fermion model with
electron and hole sectors and different kinds of fermion-boson
coupling was proposed in \cite{Batle1} to unify the BEC and BCS phenomena.
Superconductivity or superfluidity regimes take place at different
model parameters.

Despite the $d$-wave pairing symmetry is considered as a typical
property of cuprates \cite{Kirtley1}, 
there are experimental evidences that the $s$-wave or
mixed $s+d$ pairings take place in some materials at certain
values of doping. Thus, it was shown that optimally doped
$Pr_{1.855}Ce_{0.145}CuO_{4-y}$ at low temperatures
demonstrates nodeless gap inconsistent with $d$-wave pairing 
\cite{Skinta1}, and electron doped superconductor
$Sr_{0.9}La_{0.1}CuO_{2}$ demonstrate $s$-wave pairing near
optimal doping\cite{Chen1}. 
The analysis by Zhao \cite{Zhao1} shows that
dominant bulk symmetry of the order parameter in some cuprates
is extended or anysotropic s-wave symmetry.
It was also found that the crossover  
from $d$-wave pairing in underdoped and optimally
doped regime to $s+d$-wave pairing in overdoped regime takes
place in $YBa_{2}Cu_{3}O_{7-\delta}$ \cite{Yeh1}. 
Crossover from $d$-wave pairing
to $s$-wave pairing with doping near optimal $c$ was found
for electron doped $Pr_{2-x}Ce_{x}CuO_{4}$ \cite{Biswas1} and 
for $Pr_{2-x}Ce_{x}CuO_{4-y}, La_{2-x}Ce_{x}CuO_{4-y}$ \cite{Skinta2}.
However, the $s$-wave pairing is not always present in overdoped
cuprates, $d$-wave regime in overdoped $Tl_{2}Ba_{2}Cu_{3}O_{6+\delta}$
was found in \cite{Proust1}.

Below we analyze the behavior of the superconducting systems
with different inter-particle potentials in the $s$-wave and $d$-wave 
pairing channels at $T=0$ as functions of particle density
and interaction strength. We shortly review the main properties
of different systems.
We apologize for these authors whose results are not presented
here, since it is very difficult to cover all the results
in such wide and fastly growing field as theory of superconductivity,
even in a case of some special topic.

\section{\label{sec:level2} The main equations}

The most general Hamiltonians,
which is usually studied in theory of superconductivity
can be written as:
\begin{eqnarray}
H=-\sum_{n,m,\sigma }
t_{n m}c_{n\sigma}^{\dagger}c_{m\sigma}
-\mu\sum_{n,\sigma }
c_{n\sigma}^{\dagger}c_{n\sigma}
-V_{0}\sum_{n}
c_{n\uparrow}^{\dagger}c_{n\downarrow}^{\dagger}
c_{n\downarrow}c_{n\uparrow}
-\sum_{n,m}V_{nm}
c_{n\uparrow}^{\dagger}c_{m\downarrow}^{\dagger}
c_{m\downarrow}c_{n\uparrow}
\nonumber \\
+\sum_{n,q}g_{n}(q)c_{n\sigma}^{\dagger}c_{n\sigma}{\vec X}_{n}(q)+
 \frac{1}{2}\sum_{n,q}
\left[\frac{{\vec p}_{n}^{2}(q)}{m(q)}+m(q)\omega_{n}^{2}(q)
{\vec X}_{n}^{2}(q)
\right],
\label{Hamiltonianl}
\end{eqnarray}
where $c_{n\sigma}\equiv c_{\sigma}(\tau ,n)$ is a fermionic
field operator with spin $\sigma =\uparrow ,\downarrow$ at lattice site $n$
and at time $\tau$, 
$t_{ij}$ describes the nearest neighbor (n.n.),
next nearest neighbor (n.n.n.) and other hopping processes;
$\mu$ is the chemical potential of the system.
The inter-particle interaction is described by the terms
proportional to $V_{0}$ (local interaction) and
$V_{nm}$ (n.n. or n.n.n. interaction).  
The last two terms describe additional electron boson interaction
and free boson parts of the Hamiltonian, where
$q$ is a boson mode with coordinate
${\vec X}_{n}(q)$, momentum ${\vec p}_{n}(q)$ and frequency
$\omega_{n}(q)$ and $g_{n}(q)$ is the fermion-boson coupling.
One can easily pass to the continuum version of this Hamiltonian
with substitution of the n.n. hopping operator $t_{nm}$ by 
$t_{nm}\rightarrow \delta_{nm}t(1-(a^{2}/(2d)){\bf \nabla}^{2})$ (where
$a$ is the inter-cite distance, $d$ is the dimensionality
of the system), introducing radius cut off in the 
the interaction terms, etc.

In the case of the d-dimensional square lattice 
the free fermion dispersion relation in the momentum
space has the following form, when the n.n. hopping takes place 
\begin{equation}
\xi ({\bf k})=-2t\sum_{j=1}^{d}cos(ak_{j})-\mu ,
\label{xi}
\end{equation}
where ${\bf k}$ is a d-dimensional wave vector.

It is convenient to find the thermodynamic potential by using
the path integral approach for
studying the properties of a quantum many-particle system. 
This method is not necessary
in the case of the mean-field solution, but it is extremely
useful when the fluctuations are studied.

The partition function of the system is 
\begin{equation}
Z=\int D\psi^{\dagger}D\psi e^{-S}
\label{Z}
\end{equation}
with the action
\begin{equation}
S=\int_{0}^{\beta}d\tau 
\left[ 
\sum_{n,\sigma}
\psi_{n\sigma}^{\dagger}(\tau)
\partial_{\tau}
\psi_{n\sigma}(\tau)+H(\tau)
\right] .
\label{S}
\end{equation}

To study the superconducting properties of the system,
we make 
the Hubbard-Stratonovich transformation with bilocal fields
$\phi_{nm} (\tau_{1} ,\tau_{2})$ and 
$\phi_{nm}^{\dagger}(\tau_{1},\tau_{2})$
can be applied \cite{Kleinert1}:

\begin{eqnarray}
\exp \left[ \psi_{n\uparrow}^{\dagger}(\tau_{1})
\psi_{m\downarrow}^{\dagger}(\tau_{2})
V_{nm}(\tau_{1},\tau_{2})
\psi_{m\downarrow}(\tau_{2})
\psi_{n\uparrow}(\tau_{1})\right]
\nonumber \\
=\int D\phi^{\dagger}D\phi
\exp\left[ -\int_{0}^{\beta}d\tau_{1}\int_{0}^{\beta}d\tau_{2}
\left(
\frac{|\phi_{nm} (\tau_{1},\tau_{2})|^{2}}
{V_{nm}(\tau_{1},\tau_{2})}
\right.\right.
\nonumber \\
\left.\left.
-\phi_{nm}^{\dagger}(\tau_{1},\tau_{2})
\psi_{n{\downarrow}}(\tau_{1})
\psi_{m\uparrow}(\tau_{2})
-\psi_{n{\uparrow}}^{\dagger}(\tau_{1})
\psi_{m\downarrow}^{\dagger}(\tau_{2})
\phi_{nm}(\tau_{1},\tau_{2})
\right)\right] ,
\label{HS}
\end{eqnarray}
where $V_{nm}(\tau_{1},\tau_{2})$ includes the effective
inter-particle attraction due to boson coupling after intergations
over the field $X$.

Let us introduce the Nambu spinor
$$
\Psi_{n} (\tau ) = 
\left(
\begin{array}{c}
\psi_{n\uparrow} (\tau ) \\
\psi_{n\downarrow}^{\dagger} (\tau )  
\end{array}\right)\ ,
\Psi_{n}^{\dagger} (\tau ) = 
\left(
\psi_{n\uparrow}^{\dagger} (\tau ),
\psi_{n\downarrow} (\tau )  
\right) .
$$

In this case the partition function can be written as
\begin{equation}
Z=\int D\psi^{\dagger}D\psi D\phi^{\dagger}D\phi 
e^{-S(\psi^{\dagger},\psi ,\phi^{\dagger} ,\phi)},
\label{Z2}
\end{equation}
where
\begin{eqnarray}
S(\psi^{\dagger},\psi ,\phi^{\dagger} ,\phi)
\nonumber \\
=
\int_{0}^{\beta}d\tau_{1} 
\int_{0}^{\beta}d\tau_{2}
\sum_{n,m}
\{ \frac{|\phi_{nm} (\tau_{1},\tau_{2})|^{2}}
{V_{nm}(\tau_{1},\tau_{2})}
-\delta (\tau_{1}-\tau_{2})
\Psi_{n}^{\dagger}(\tau_{1})
\left[-\delta_{nm}\partial_{\tau_{2}}-{\hat \tau}_{z}(t_{nm}
-\delta_{nm}\mu)\right]
\Psi_{m}(\tau_{2})
\nonumber \\
-
\phi_{nm}^{\dagger}(\tau_{1},\tau_{2})
\Psi_{n}^{\dagger}(\tau_{1})
{\hat \tau}_{-}
\Psi_{m}(\tau_{2})
-\Psi_{n}^{\dagger}(\tau_{1})
{\hat \tau}_{+}
\Psi_{m}(\tau_{2})
\phi_{nm}(\tau_{1},\tau_{2})\} ,
\label{Sb}
\end{eqnarray}
where ${\hat \tau}_{\pm}=\frac{1}{2}({\hat \tau}_{x}\pm {\hat \tau}_{y})$ 
are the Pauli matrices.

The last action is diagonal on fermionic fields,
therefore the integration over $\Psi^{\dagger}$ and $\Psi$ can be performed
exactly. The partition function becomes in this case
$$
Z
= 
\int{\cal D}\phi {\cal 
D}\phi^{*}\exp(-\beta\Omega [{\cal G}]), \ \ (\beta =1/T) 
$$ 
where $\Omega [{\cal G}]$ is the thermodynamic potential, which in the 
"leading order" is 
\begin{equation} 
\beta\Omega [{\cal G}]=
\int_{0}^{\beta}d\tau_{1}
\int_{0}^{\beta}d\tau_{2}
 \sum_{n,m}
\frac{|\phi_{nm} (\tau_{1},\tau_{2})|^{2}}
{V_{nm}(\tau_{1},\tau_{2})}
- Tr Ln G^{-1}
+ Tr Ln G_{0}^{-1}.
\label{effpot} 
\end{equation}
The Nambu spinor Green function $G$ satisfies the following
equation
\begin{eqnarray}
\left[-\delta (\tau_{1}-\tau_{3})\partial_{\tau_{3}}{\hat I}+
\delta (\tau_{1}-\tau_{3})(t_{nm}-\delta_{nl}\mu ){\hat \tau}_{z}
+{\hat \tau}_{+}\Phi_{nl} (\tau_{1},\tau_{3})
+{\hat \tau}_{-}\Phi_{nl}^{*}(\tau_{1},\tau_{3}) 
\right]G_{lm}(\tau_{3} ,\tau_{2})
\nonumber \\
=\delta (\tau_{1}-\tau_{2} )\delta_{nm} 
\nonumber
\end{eqnarray}
with anti-periodic boundary conditions
$$
G_{nm}(\tau_{1}-\tau_{2} +\beta)=-G_{nm}(\tau_{1}-\tau_{2}).
$$
The thermodynamic potential (\ref{effpot}) is the most
general form of the superconducting effective potential with non-local
retarded inter-particle interaction. It will be used below to study the
fluctuation effects.

The minimization of the thermodynamic potential with
respect to the order parameter and chemical potential 
leads to the
following system of the coupled equations for $\Phi_{nm} (\tau_{n},\tau_{m})$
and $\mu$:
\begin{equation}
\frac{\delta \Omega}{\delta \Phi_{nm} (\tau_{n},\tau_{m})}=0
\label{gap}
\end{equation}
\begin{equation}
\frac{\partial \Omega}{\partial \mu}=-N_{f},
\label{mu}
\end{equation}
or
\begin{equation}
\Phi_{nm} (\tau_{1},\tau_{2})=
V_{nl} (\tau_{1},\tau_{3})tr{\hat \tau}_{+}G_{lm}(\tau_{3},\tau_{2})
\label{gap2}
\end{equation}
\begin{equation}
n_{f}=-tr{\hat \tau}_{z}G_{nn}(\tau_{1},\tau_{1}^{+}),
\label{mu2}
\end{equation}
where $n_f=N_{f}/v$ is particle density in the system
($v$ is the volume of the system).

In general it is very difficult to find the Greens function
$G_{nm}(\tau_{1},\tau_{2})$, therefore some simplifications must be applied.
We shall consider the case of the space and time invariance,
$V_{nm}(\tau_{1},\tau_{2})=V_{n-m,0}(\tau_{1}-\tau_{2})$.
In this case the Green's function has the following form in the
momentum space: 
$$
G(i\omega_{n},{\bf k})
=-\frac{i\omega_n +{\hat \tau}_{3}\xi ({\bf k})
        -\Phi (i\omega_{n},{\bf k}){\hat \tau}_{+}
        -\Phi^{*}(i\omega_{n},{\bf k}){\hat \tau}_{-}}
{\omega_n^{2} +\xi^{2} ({\bf k})+|\Phi (i\omega_{n},{\bf k})|^{2}},
$$
and the system of the equations 
(\ref{gap2}), (\ref{mu2}) has the following form:
\begin{eqnarray}
\Phi (i\omega_{n},{\bf k})=
\int\frac{d^dp}{(2\pi )^{d}}\sum_{m}
\frac{\Phi (\omega_{m},{\bf p})}{\omega_{m}^{2}-\xi ({\bf p})^{2}
-|\Phi(\omega_{m},{\bf p})|^{2}}
\nonumber \\
\times\left[
V({\bf p},{\bf k})
+g_{b}^{2}\int\frac{d^dk}{(2\pi )^{d}}
\frac{\omega ({\bf p}-{\bf k})^{2}}
{-(\omega_{m}-\omega_{n})^{2}-\omega ({\bf p}-{\bf k})^{2}}
\right]
\label{gap3}
\end{eqnarray}
\begin{equation}
n_{f}=\int\frac{d^dp}{(2\pi )^{d}}
\left[ 1-
\sum_{m}
\frac{\xi ({\bf p})}{-\omega_{m}^{2}-\xi ({\bf p})^{2}
-|\Phi (\omega_{m},{\bf p})|^{2}}
\right] ,
\label{mu3}
\end{equation}
$V({\bf p}-{\bf k})$ is the Fourier transform of the
non-retarded interaction and the term proportional
to $g_{b}$ describes the inter-partice boson coupling,
$\omega_{n}=\pi T(2n+1)$ is the standard Matsubara frequency.
Interaction term in (\ref{gap3}) is written in general form.
it describes, for example local non-retarded interaction,
when $V({\bf p}-{\bf k})=const, g_{b}=0$, non-local non-retarded
interaction, when $V({\bf p}-{\bf k})\not= const, g_{b}=0$,
local retarded interaction, when $V({\bf p}-{\bf k})=const, g_{b}\not= 0,
\omega ({\bf p}-{\bf k})=const$ etc.
The system (\ref{gap3}), (\ref{mu3}) will be analyzed in the next
Section for different forms of the inter-particle potential
V({\bf p},{\bf k}) and spectra $\omega ({\bf p}-{\bf k})$.

\section{Solutions}

\subsection{Model with local non-retarded attraction}

The problem of the crossover from small to large fermion density
in the model with local attraction
was considered in \cite{Eagles,SadeMelo1,Haussmann1,Randeria3,Micnas1,Engelbrecht1,Marini1,Andrenacci1,Babaev1} for the 3D case
and in \cite{Eagles,Randeria1,Micnas1,Andrenacci1,Gorbar2} for the 2D case.

For the simplest case of local non-retarded attraction
the interaction parameters
in equations (\ref{gap3}), (\ref{mu3}) have the following form:
$V({\bf p},{\bf k})\equiv V=const, g_{b}=0$. Therefore, the gap
in this case is momentum and frequency independent
$$
\Phi (\omega_{n},{\bf k})=\Delta =const.
$$
The summation over frequency in (\ref{gap3}), (\ref{mu3}) 
can be easily performed,
and one gets the standard system of the equations:
\begin{equation}
1=V
\int\frac{d^dk}{(2\pi )^{d}}
\frac{1}{2\sqrt{\xi ({\bf k})^{2}+\Delta^{2}}}
\tanh\left( \frac{\sqrt{\xi ({\bf k})^{2}+\Delta^{2}}}{2T}
\right) ,
\label{gaplocal}
\end{equation}
\begin{equation}
n_{f}=\int\frac{d^dk}{(2\pi )^{d}}
\left[ 1-
\frac{\xi ({\bf k})}{\sqrt{\xi ({\bf k})^{2}+\Delta^{2}}}
\tanh\left( \frac{\sqrt{\xi ({\bf k})^{2}+\Delta^{2}}}{2T}
\right)
\right],
\label{mulocal}
\end{equation}
which have a simple form at $T=0$:
\begin{equation}
1=V
\int\frac{d^dk}{(2\pi )^{d}}
\frac{1}{2\sqrt{\xi ({\bf k})^{2}+\Delta^{2}}}
\label{gaplocalT0}
\end{equation}
\begin{equation}
n_{f}=\int\frac{d^dk}{(2\pi )^{d}}
\left[1-
\frac{\xi ({\bf k})}{\sqrt{\xi ({\bf k})^{2}+\Delta^{2}}}
\right] .
\label{mulocalT0}
\end{equation}
Since the gap is momentum independent, the only isotropic $s$-wave pairing
regime is possible in the model with local attraction.

The $s$-wave pairing regime in the case of quadratic dispersion
law
$$
\xi ({\bf k})=\frac{{\bf k}^{2}}{2m}-\mu
$$
and indirect pairing with boson energy 
cut-off $\theta (|\xi ({\bf p})-\mu | -\omega_{c})$
was considered in the 2D and 3D case for low carrier densities
by Eagles \cite{Eagles}.
In fact, in this case the integration over momentum 
$\int d^dk/(2\pi )^{d}$ can be replaced
by integration over energy $\int \rho (\epsilon )d\epsilon$, where
$\rho (\epsilon )$ is the density of states (DOS), 
it is constant in the 2D case
and $\sim \sqrt{\epsilon}$ in the 3D case. 

It is very easy to solve system (\ref{gaplocalT0})
and (\ref{mulocalT0}) in the 2D case. In the case when 
$\epsilon_{F}<<W$ ($W$ is a free fermion bandwidth) 
the solution has a simple form:
$$
\Delta \simeq\sqrt{2W\epsilon_{F}}e^{-2\pi/(mV)},
$$
$$
\mu \simeq \epsilon_{F}-\frac{|\varepsilon_{b}|}{2},
$$
where
$\varepsilon_{b}=-2We^{-4\pi/(mV)}$ is the two-particle 
bound state energy.
Obviously, the crossover from superfluidity to superconductivity
with doping takes place at any coupling constant, there 
exists the value of $\epsilon_{F}$ when $\mu =0$
for any $V$.

This is not true in the 3D case, when the crossover takes
place only when the coupling constant is larger than some critical value
$V_{c}$. This difference follows from the difference in the DOS
 in the gap equation.
In the 2D case $\rho (\epsilon )=const$ and the gap equation
has the solution $\Delta =\sqrt{2W|\varepsilon_{b}|}$
at $\mu =0$ and any coupling constant.

In the 3D case the gap equation has the following form
\begin{equation}
1=V
\int\frac{k^2dk}{2\pi^{2}}
\frac{1}{2\sqrt{k^{4}+\Delta^{2}}}
\label{gapT03D}
\end{equation}
at $\mu =0$.
The integral over $k$ on the right hand side
has maximal value $\sqrt{2mW}/(4\pi^{2})$ at $\Delta =0$.
Therefore, a simple estimation for $V_{c}$ is given by the
relation $1\simeq V_{c}\sqrt{2mW}/(4\pi^{2})$, or 
$V_{c}\simeq 4\pi^{2}/\sqrt{2mW}$.

In the case of the momentum cut-off
$\int\frac{d^dk}{(2\pi )^{d}}\theta (|\xi ({\bf p})-\mu | -\omega_{c})$
the approximate solution can be also easily obtained in the 2D case:
$$
\Delta \simeq\sqrt{2|\varepsilon_{b}|\Delta_{BCS}}\theta (\omega_{c}-\epsilon_{F})
+\Delta_{BCS}\theta (\epsilon_{F}-\omega_{c})
$$
$$
\mu = \epsilon_{F}-\frac{|\varepsilon_{b}|}{2},
$$
where $\Delta_{BCS}=2\omega_{c}e^{-2\pi/(mV)}$ is the BCS expression
for the gap and the bound state energy in this case
is $\varepsilon_{b}=-2\omega_{c}e^{-4\pi/(mV)}$. 
The gap is growing with doping in this case, and asymptotically
reaches its maximal
value $\Delta=\Delta_{BCS}$ when $\epsilon_{F}>>\omega_{c}$.

In the 3D case the DOS in equations (\ref{gaplocalT0}) 
and (\ref{mulocalT0}) can be substituted
by the DOS on the Fermi level,
and the 3D solution can be obtained from the 2D one with substitution
$mV/(2\pi )\rightarrow k_FmV/(4\pi^{2})$.
In is possible to estimate the critical value of the coupling constant
when the crossover takes place: $V_{c}=2\pi^{2}/\sqrt{m\omega_{c}}$.
In both cases with and without momentum cut off the transition
from superfluidity to superconductivity is smooth, in other words the gap 
is groving with doping continuously.

\subsection{Models with non-local non-retarded interaction}

It is important to study a more realistic case of the non-local attraction
in presence of the short range Coulomb repulsion.
In order to study the superconducting properties of such a model
in the channels with different pair angular momentum $l$,
it is convenient (see  \cite{Borkowski1,Duncan1,magnons}, for example), 
to approximate the interaction potential by  
by a separable function:
\begin{equation}
V_{l}({\bf k}_1,{\bf k}_2)=-\lambda_{l}w_{l}({\bf k}_1)w_{l}({\bf k}_2),
\label{potentialk}
\end{equation}
where $\lambda_l$ is an effective coupling constant, and 
\begin{equation}
w_{l}({\bf k})=h_{l}(k)cos(l\varphi_{\bf k}),
\label{w}
\end{equation}
\begin{equation}
h_{l}(k)=\frac{(k/k_1)^l}{(1+k/k_0)^{l+1/2}},
\label{h}
\end{equation}
$k$ is the momentum modulus $k=|{\bf k}|$ and $\varphi_{\bf k}$ is 
the momentum  
angle in polar coordinates
${\bf k}=k( cos(\varphi_{\bf k}), sin(\varphi_{\bf k}))$.
Parameters $k_0$ and $k_1$ put the momentum range in the proper
region - the potential is attractive at $r_{0}<r<r_{1}$
and repulsive at $r<r_{0}$, where $k_0\sim 1/r_0$ and $k_1\sim 1/r_1$.

It is easy to see  that the interaction potential (\ref{potentialk})
 has the correct asymptotic behavior at small and large 
momenta: $V_{l}({\bf k}_1{\bf k}_2)\sim k_1^lk_2^l$ and 
$V_{l}({\bf k}_1{\bf k}_2)\sim 1/\sqrt{k_1k_2}$, correspondingly.
Since the region of low carrier concentrations,
where the crossover can take place, is the most interesting,
the correct behavior of the
interaction potential at small momenta is the most important.
The small momenta give the main contribution to the
integrals in the case of low carrier  concentrations
(see equations below).
We shall study $s$- and $d$-wave channels with $l=0$ and $2$ separately, so
we assume that the parameters $\lambda_l$ for both channels
are independent.

In this case the equations for the gap and for the chemical
potential have the following form:
\begin{equation}
\Delta_{l} ({\bf k}) =-\lambda_{l}\int\frac{d {\bf p}}{(2\pi)^{2}}
\frac{\Delta_{l} ({\bf p})}{2\sqrt{\varepsilon^{2}({\bf p})+\Delta_{l}({\bf p})^{2}}}
V_{l}({\bf p},{\bf k}), 
\label{gapequation}
\end{equation}
\begin{equation}
\int\frac{d {\bf p}}{(2\pi)^{2}}
\left[ 1-\frac{\varepsilon ({\bf p})}{\sqrt{\varepsilon^{2}({\bf p})+\Delta_{l}({\bf p})^{2}}}
\right] 
=n_{f}.
\label{numberequation}
\end{equation}
The solution of  equation (\ref{gapequation}) has the following form
\begin{equation}
\Delta_{l} ({\bf k}) =\Delta_{l}^{(0)} w_{l}({\bf k}),
\label{gap}
\end{equation}
where $\Delta_{l}^{(0)}$ does not depend on the momentum ${\bf k}$.

As it was shown in \cite{Borkowski1,Duncan1}, the crossover
from superfluidity to superconductivity is smooth
with doping increasing (see Figs. 1 and 2 below).
However, there exists a critical value of the
interaction potential in the $d$-wave pairing channel,
below which the crossover from superconductivity to superfluidity
is impossible in the $d$-wave pairing channel.

The more realistic case in connection to HTSCs was considered in 
\cite{magnons}, where the correlation length $r_{0}$
was considered at small carrier densities
as $r_0\sim a/\sqrt{n_f}$. This dependence
for the length of spin-spin correlations
at small carrier densities was found for example in
 $La_{2-x}Sr_{x}CuO_{4}$.
The magnetic correlation length decreases with carrier density per cell
in this material as $3.8$\AA$/\sqrt{n_f}$ \cite{Thurston}.
The value of $a$ was considered to be 
$a =\sqrt{2/\pi}a_0$, $a_0$ is the square lattice constant.
The following equality gives the value for $a$:
$(\pi /2) r_{0}^{2}N_{f}=a_{0}^{2}N_{cell}$,
where on the left side the volume of the 2D system is expressed as a
volume (circle of the radius $\sim r_{0}$) occupied by one particle,
multiplied by the full number of particles $N_{f}$, 
$N_{cell}$ is an elementary cell number in the system.
The free fermion bandwidth $W$ is connected with $a_{0}$ as
$W=\pi^{2}/(ma_{0}^{2})$.
It should be noted, that the relation $r_0\sim a/\sqrt{n_f}$
at $a=\sqrt{2/\pi}a_{0}$ 
is in a good agreement with the experimental data for
$La_{2-x}Sr_{x}CuO_{4}$ \cite{Thurston}, where the plane
magnetically ordered lattice parameters are equal 
to $5.354$ \AA \ \ and $5.401$\ \ \AA , 
and the corresponding parameter $a$  is $\simeq 3.8$ \AA.

It was shown the the critical value of the coupling exists
even in the $s$-wave pairing channel for this case (see Figs.1 and 2). 

\begin{figure}[h]
\centering{
\includegraphics[width=8.0cm,angle=270]{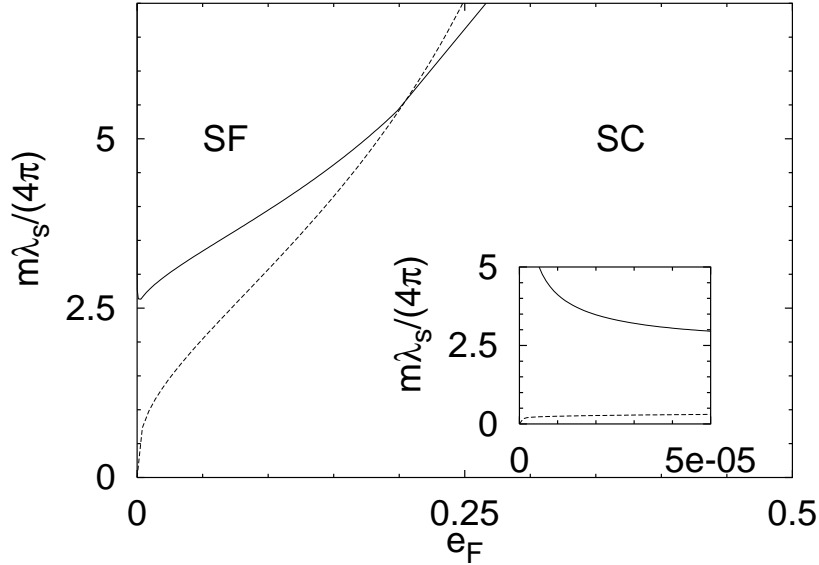}}
\caption{
Crossover line coupling-carrier density 
is presented for the $s$-pairing channel
(solid line).
 The dotted line represents the corresponding curve
for the case $r_{0}(n_{f})=const$ at $r_{0}=a_{0}$.
The insert shows the doping dependence of the crossover
value for coupling at very low charge carrier densities
\cite{magnons}.
Here and below all parameters are expressed in units of the bandwidth $W$
}  
\label{fig:1}
\end{figure}

\begin{figure}[h]
\centering{
\includegraphics[width=8cm,angle=270]{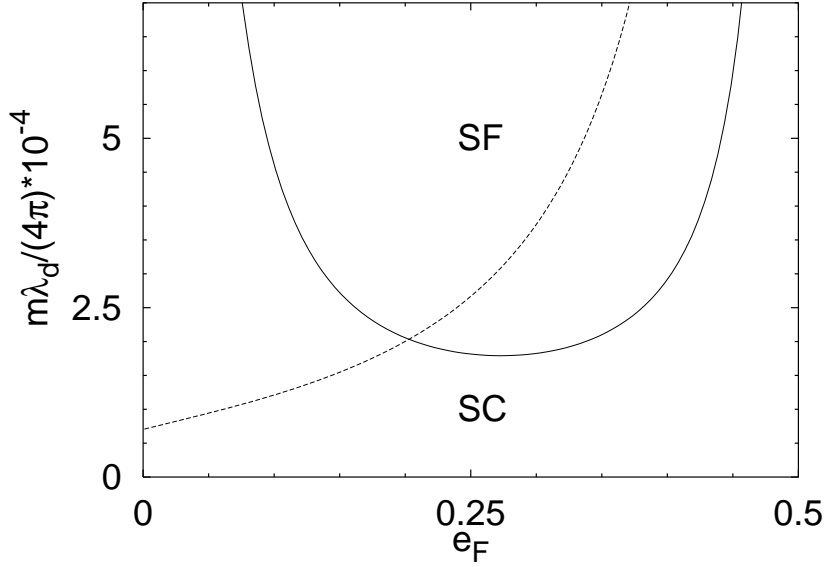}}
\caption{The crossover line coupling-carrier density
is presented for the $d$-wave case (solid line). 
The dotted line is the
crossover curve for the case $r_{0}(n_{f})=const$ at $r_{0}=a_{0}$
\cite{magnons}.}  
\label{fig:2}
\end{figure}

Different versions of the previous model can be considered.
Namely, the correlation radius 
$r_{0} (n_{f})= a_{0}\sqrt{2/(\pi n_f)}$ can be introduced 
in a model with exponential decay of the attraction:
\begin{equation}
V_{b}(r,t)=g_{e-b}\frac{e^{-r/r_{0}}}{r}
\label{magnoninteraction}
\end{equation}
Its Fourier transform has the following form:
\begin{equation}
D_{b}({\bf q})=\frac{g_{e-b}r_{0}}{(2\pi)^{2}\sqrt{1+r_{0}^{2}{\bf q}^{2}}}.
\label{magnoninteractionp}
\end{equation}
In addition, the same kind of the short-range repulsion
can be introduced in the model. It is easy to see, that in the 2D
case this potential has the following form (see, for example \cite{Fetter}):
\begin{equation}
D_{e}({\bf q})=\frac{g_{e-e}}{q+q_{TF}g(q/2k_{F})},
\label{Coulombinteraction}
\end{equation}
where $g_{e-e}\equiv 2\pi e^{2}$, the Thomas-Fermi momentum is 
$q_{TF}=4e^{2}m/\pi =4/\pi a_{B}$, and
$a_{B}=1/(e^{2}m)\simeq 0.529$ \AA is the Bohr radius.
The function $g(x)$ is defined as
\begin{equation}
g(x)=1-\theta (x-1)\sqrt{1-1/x^{2}}.
\label{g}  
\end{equation}

This model in the $s$-wave pairing channel demonstrates
the crossover from superfluidity to superconductivity
at any value of the coupling constant, contrary
to the previous case, see Fig.3.
The doping dependence of the gap and of the chemical potential
at different values of the dimensionless interaction
parameters
$\lambda_{e-b}=g_{e-b}^{2} mr_{0}/(8\pi^{2}), \ \ \ 
\lambda_{e-e}=g_{e-e}^{2} mr_{0}/(4\pi )$
is presented in Figs. 4 and 5.
As it follows from these Figures, the gap is decreasing with
doping growing. This situation is in a qualitative agreement
with the experiments for the cuprates. 
\begin{figure}[h]
\centering{
\includegraphics[width=8.0cm,angle=270]{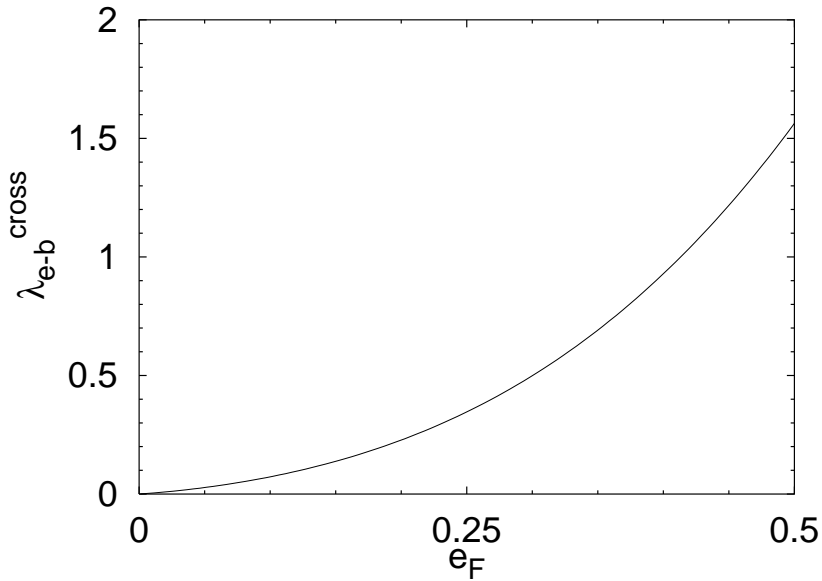}}
\caption{The crossover interaction value 
$\lambda_{e-b}^{cross}$ (defined by the condition $\mu =0)$ is presented
as a function of $\epsilon_{F}$ at $\lambda_{e-e}=0$.
}  
\label{fig:3}
\end{figure}
\begin{figure}[h]
\centering{
\includegraphics[width=5.5cm,angle=270]{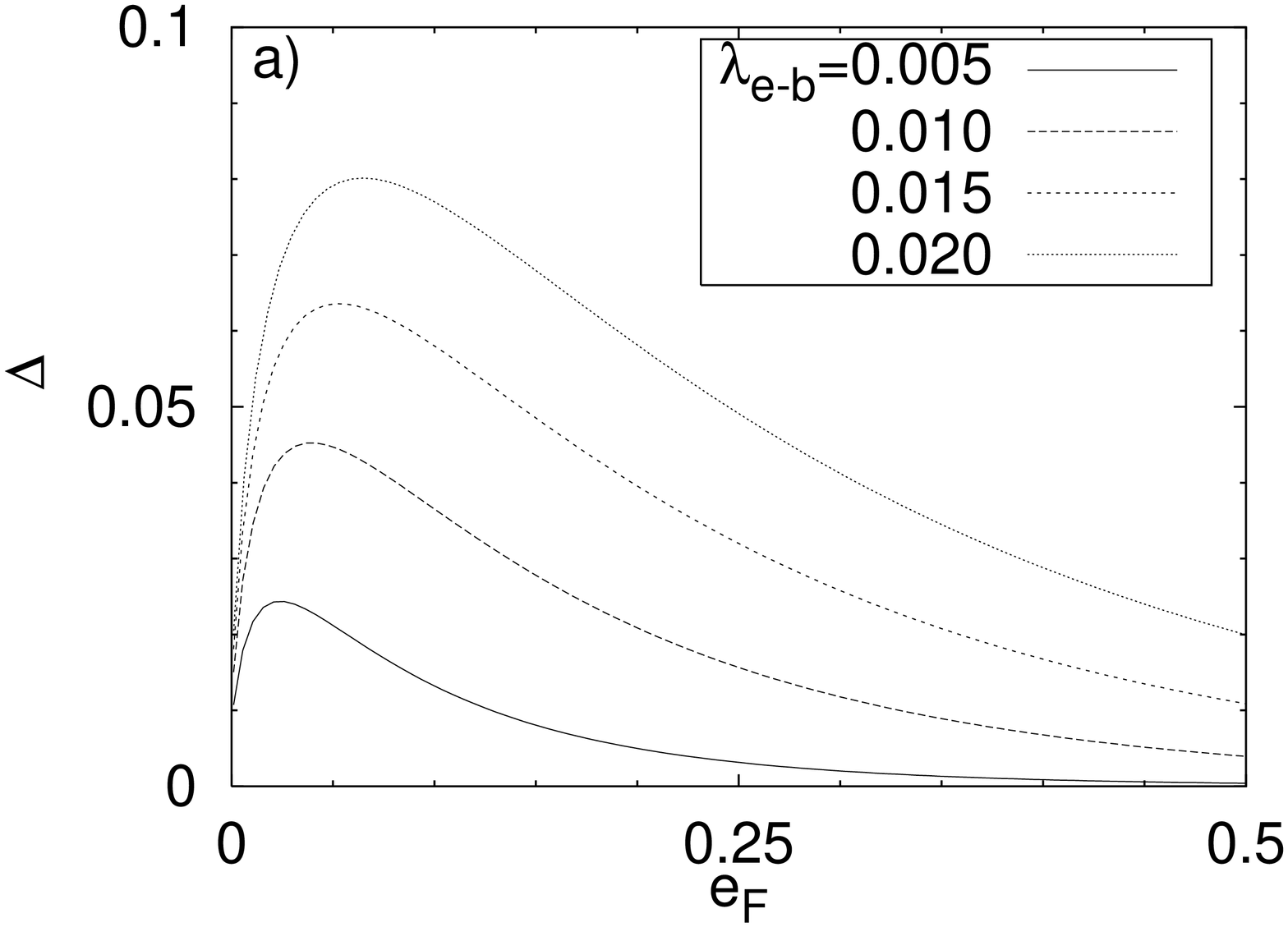}
\includegraphics[width=5.5cm,angle=270]{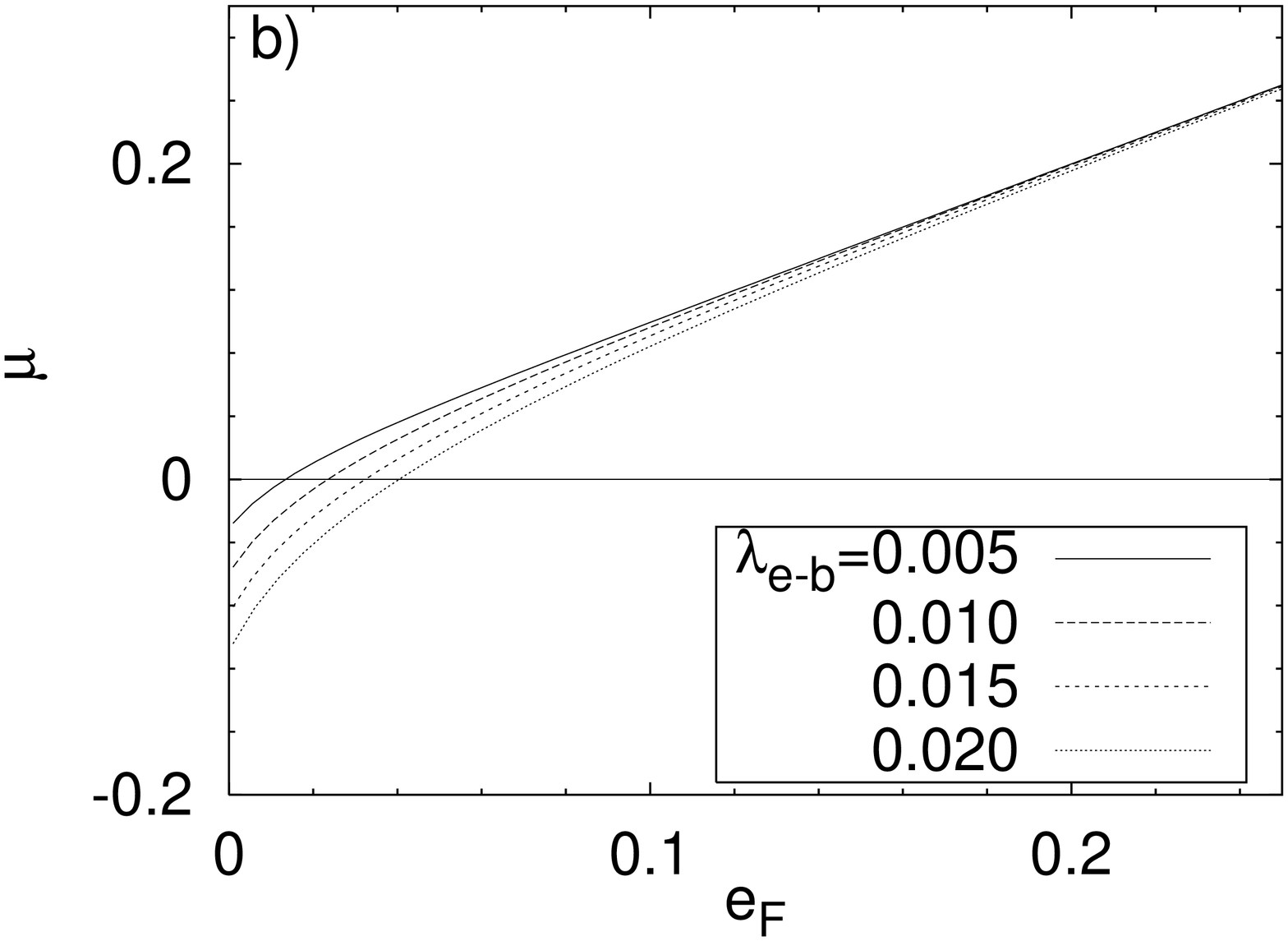}}
\caption{
The gap (a) and the chemical potential (b) as functions of $\epsilon_{F}$ 
at different $\lambda_{e-b}$ and $\lambda_{e-e}=0$.
}  
\label{fig:4}
\end{figure}
\begin{figure}[h]
\centering{
\includegraphics[width=5.5cm,angle=270]{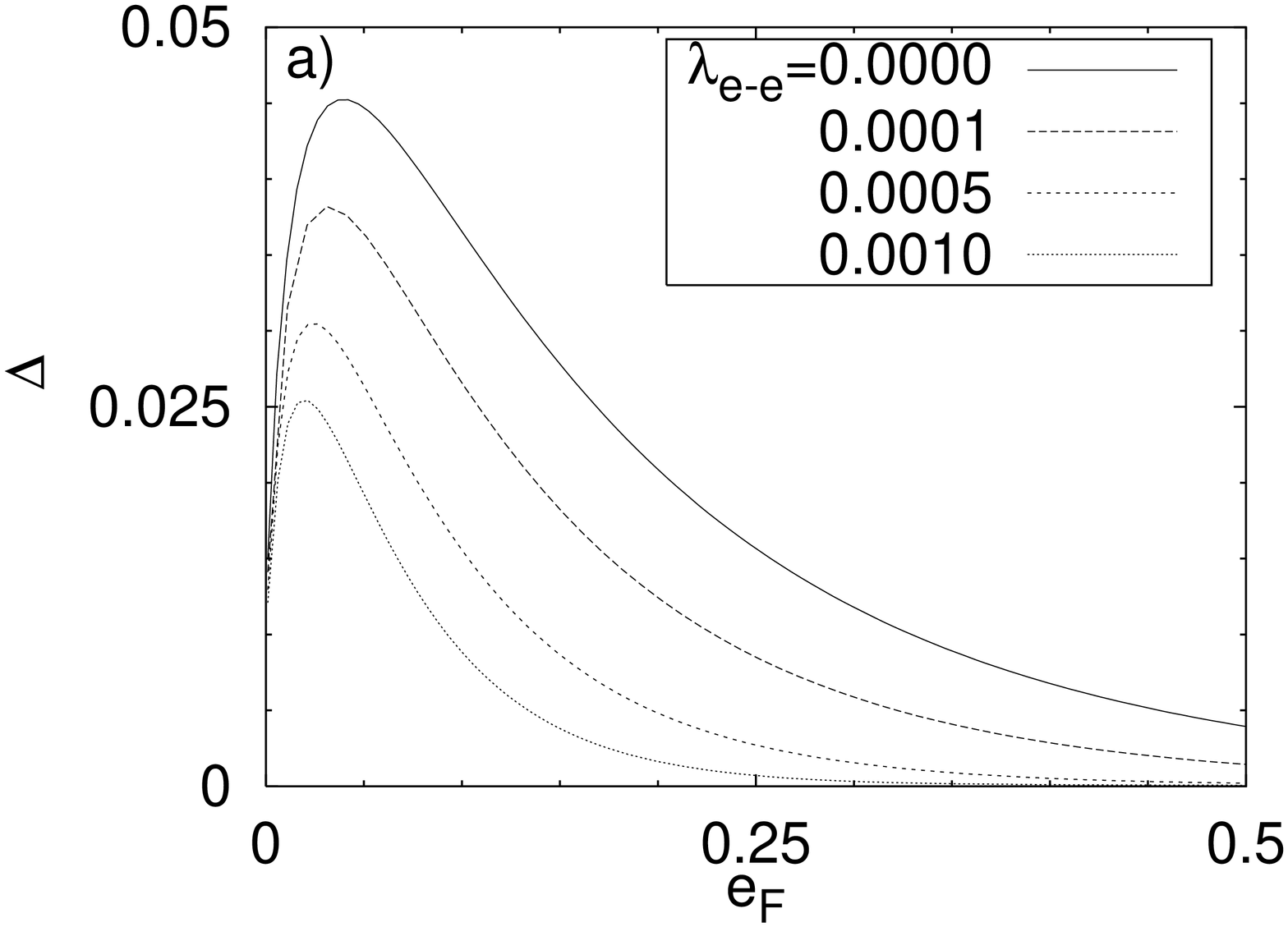}
\includegraphics[width=5.5cm,angle=270]{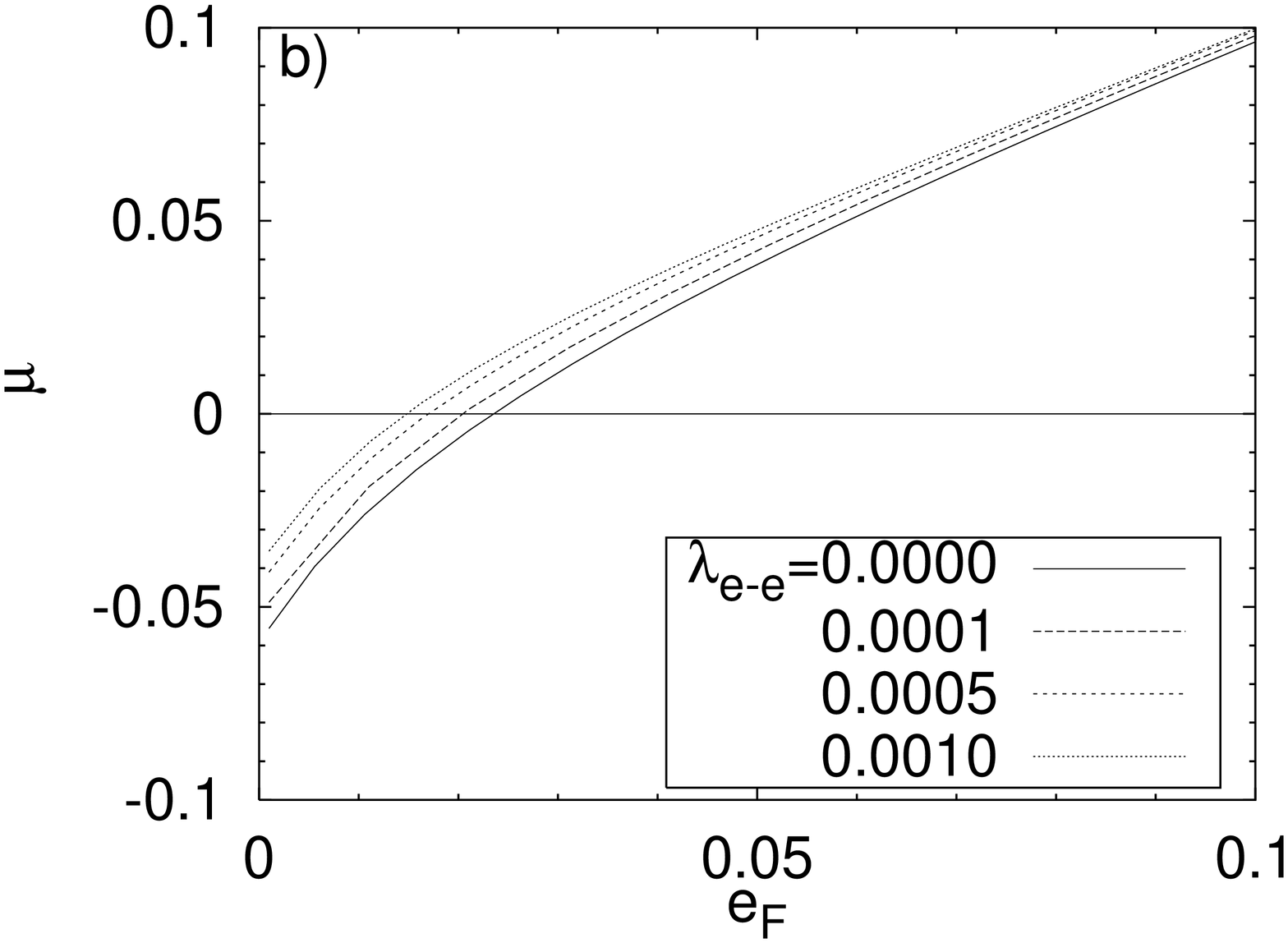}}
\caption{The gap (a) and the chemical potential (b) 
as functions of $\epsilon_{F}$ 
at different $\lambda_{e-e}$ and $\lambda_{e-b}=0.01$.}  
\label{fig:5}
\end{figure}
One can consider also an interesting situation of the combined
local+non-local attraction, when the local attraction will
tend the Cooper pairs to transform into the local pairs. 

The ``mixed'' case with $h_{l} (k)=1$ and 
$\lambda =\lambda_{s}+\lambda_{d}cos(2\phi_{\bf k})$
was considered in Ref. \cite{Musaelian1,Gorbar3}.
In particular, it was shown in \cite{Musaelian1} 
that the crossover from the $d$-wave 
to the $s$-wave superconductivity with the intermediate
$s+d$-phase takes place with doping
in agreement with the experiments
on some cuprate materials
\cite{Yeh1,Biswas1,Skinta2}.

The $d$-wave case $w_{d}({\bf k})=cos(k_{x})-cos(k_{y})$
was considered in \cite{denHertog1,Soares1,Perali1}.
In particular, the role of the next nearest neighbor hopping
$t'$ on the on the pairing was studied in \cite{Soares1,Perali1}.
As it was shown in \cite{Perali1}, the crossover with doping
take place at any value of $t'$, when the coupling constant
 is bigger than $V_{c}$.

It was stated in \cite{Fehrenbacher1}, that
the model with small on-site repulsion
and nearest neighbor $V_{1}$ and next nearest neighbor $V_{2}$
attraction with $V_{2}\sim 60-80$meV and $V_{2}/V_{1}\simeq 1.3-1.5$
can describe the experimental data for hole-doped oxides.

\subsection{Models with retarded interaction}

The retardation effects in the interaction can play very important
role the superconducting properties of the system.
Let us consider a boson propagator with the dispersion
$\omega = \omega ({\bf k})$:
\begin{equation}
D(\omega, {\bf k})=\frac{\omega ^{2}({\bf k})}
{\omega^{2}- \omega^{2}({\bf k})+i\delta}, 
\label{Prop} 
\end{equation}
where the spectra $\omega ({\bf k})$.
For example, in the general case of phonon dispersion one has
$\omega ({\bf k})=\sqrt{\omega_{0}^{2}+c^{2}{\bf k}^{2}}$.
We shall study this rather general case below with the following
approximation:
$D (\omega ,{\bf k})\simeq D (\omega ,{\bf k}_F)$,
so the effective propagator can be written as
\begin{equation}
D(\omega, {\bf k})=\frac{\omega_{0}^{2}}{\omega^{2}- \omega_{0}^{2}+i\delta},
\label{Prop2} 
\end{equation}
where $\sqrt{\omega_{0}^{2}+c^{2}{\bf k}_{F}^{2}}$ is substituted
by new effective frequency $\omega_{0}$.
This approximation corresponds to the case of the optical phonon
attraction (in the case when $\omega_{0}\not= 0$). 
Generally speaking, the gap is frequency-dependent in
this model. The problem of the crossover in 
the model with frequency dependent gap was studied in \cite{TMF}.

The system of the equations for the gap and for the
chemical potential in this case is
\begin{equation} 
\phi(\omega )=-ig^{2}\int\frac{d^{2}kd\nu}{(2\pi)^{2}}
\frac{\phi (\nu )}
{\nu^{2}-\xi (\mbox{\bf k})^{2}-|\phi (\nu )|^{2}+i\delta}
\frac{\omega_0 ^2}
{(\omega -\nu )^{2}-\omega_{0}^{2}+i\delta},\label{Gap2}
\end{equation}
\begin{eqnarray} 
2\epsilon_{F}=Re\int_{0}^{\infty}d\omega[(\frac{\omega}{\sqrt{\omega^{2}-
|\phi (\omega)|^{2}}}-1)\theta(\mu +\sqrt{\omega^{2}-|\phi (\omega)|^{2}})+
                                                         \nonumber \\ 
(\frac{\omega}{\sqrt{\omega^{2}-|\phi 
(\omega)|^{2}}}+1)\theta (\mu - \sqrt{\omega ^{2}-|\phi 
(\omega)|^{2}})].
\label{Num1} 
\end{eqnarray}

Let us describe shortly how this system of the equations can be analyzed
analytically.
First of all, it is possible to show that the approximation
$\phi (\omega )=\Delta =$ const in (\ref{Num1}) is a rather good one.
Then, this equation gives:
$$
\mu =\epsilon_{F}-\frac{\Delta^{2}}{2\epsilon_{F}}\simeq 
\epsilon_{F}-\frac{|\varepsilon_{b}|}{2},
$$ 
where the two-particle bound state energy $\varepsilon_{b}$ 
depends on the coupling parameter in this case (see below).

After the Wick rotation
$\omega\rightarrow i\omega$ the gap equation becomes
\begin{equation} 
\phi (\omega )=g^2\omega_0^2\int\frac{d^{2}kd\nu }{(2\pi)^3}\frac{\phi (
\nu )} 
{\nu^2+({\bf k}^2 /2m-\mu)^2+\phi^2(\nu )}
\frac{1}{(\omega -\nu )^2+ 
\omega_0^2},\label{E11}
\end{equation}
In polar coordinates, after integration
over the angle we obtain:
$$ 
\phi (\omega )=\frac{g^2\omega_0^2}{(2\pi)^2}\int_{-\infty}^{+\infty}d\nu 
\frac{\phi (\nu )}{(\omega -\nu )^2+\omega_0^2}
\int_{0}^{\infty}\frac{kdk} 
{(k^2 / 2m - \mu)^2+\nu^2+\phi^2(\nu )}.
$$

Since $\phi (\omega)$ is an even function of $\omega$, it can depend
only on $\omega^{2}$, we restrict the integration
over $\nu$ to the positive values:

\begin{equation} 
\phi (\omega )=\frac{\lambda\omega_0^2}{2}\int_{0}^{\infty}\frac{d\nu\phi 
(\nu )} 
{\sqrt{\nu^2+\phi^2(\nu )}}\frac{1}{(\omega -\nu )^2+\omega_0^2}
\left[\frac{\pi}{2}+
\arctan\frac{\mu}{\sqrt{\nu^2+\phi^2(\nu )}}\right].
\label{E12}
\end{equation}
The coupling constant $\lambda =g^{2}m/(2\pi )$ is introduced.
The asymptotic behavior for $\phi (\omega )$ is

\begin{equation} 
\phi (\omega )\mid_{\omega\rightarrow const}\rightarrow 0,\qquad 
\phi (\omega )\mid_{\omega\rightarrow \infty}\sim 
\frac{1}{\omega^2}.\label{E12a} 
\end{equation}

We make the next replacement in the interaction 
potential \cite{Appelquist1,Gusynin1}:
\begin{equation} 
\frac{1}{(\omega -\nu )^2+\omega_0^2}
=\frac{1}{\omega^2+\omega_0^2}\theta (\omega -\nu )+
\frac{1}{\nu^2+\omega_0^2}\theta (\nu -\omega ).\label{E12b}
\end{equation}

Then, the differentiation with respect to $\omega$ gives:
\begin{equation} 
\phi '(\omega )=-\frac{\lambda\omega_0^2\nu}{(\nu^2+\omega_0^2)^2} 
\int_{0}^{\omega}
\frac{d\nu\phi (\nu )}{\sqrt{\nu^2+\phi^2(\nu )}}
\left[1+
\frac{2}{\pi}\arctan\frac{\mu}{\sqrt{\nu^2+\phi 
^2(\nu )}}\right],
\label{E14} 
\end{equation}
It is evident that $\phi '<0$, i.e. $\phi_{max}=\phi (0)\equiv \Delta$.

Differentiating once more and introducing a new variable 
$x =\nu^{2}/\omega_{0}^{2}$ one gets the following differential
equation 
\begin{eqnarray} 
\phi''(x)+\frac{2}{x+1}
\phi'(x)+\frac{\lambda}
{4\sqrt{x}(x+1)^{2}\sqrt{x+(\phi (x)/\omega_0)^2}}    
\nonumber \\
\times
\left[1+                                        
\frac{2}{\pi}\arctan\frac{\mu/\omega_0}
{\sqrt{ x + (\phi (x)/\omega_0)^{2} }}\right]\phi (x)=0.
\label{Gap3} 
\end{eqnarray}
with the boundary conditions
\begin{equation} 
\left.\phi '(x) \right|_{{x}=0}=0;\qquad
\left. \left [\phi (x) 
+ (x+1)\phi '(x)\right] 
\right|_{x=\infty}=0, \label{Gap3bc}
\end{equation}
These conditions follow directly 
from (\ref{E12}), (\ref{E12b}) and (\ref{E14}).

The analysis of this equation shows that the approximate
solution for the gap in both weak and strong coupling regimes
is (see \cite{TMF} for details)
$$
\Delta (\omega)\simeq \Delta\theta (\omega_{0}^{2}-\omega^{2}),
$$
where $\Delta$ is a parameter, which 
depends on the coupling and the carrier density, what is very important.
In the weak coupling regime the parameter
$\Delta$ has the following coupling and carrier density dependence: 
$$
\Delta =\sqrt{2|\varepsilon_{b}|\omega_{0}}\theta (\omega_{0}-\epsilon_{F})
+\Delta_{BCS}\theta (\epsilon_{F}-\omega_{0})
$$
where $\Delta_{BCS}=2\omega_{0}e^{-1/\lambda}$ is the BCS expression
for the gap and the bound state energy in this case
is $\varepsilon_{b}=-2\omega_{0}e^{-2/\lambda}$.
In the strong coupling regime $\varepsilon_{b}\simeq \lambda$ and
$\Delta_{BCS}\rightarrow (4/3)\lambda\omega_{0}$ at large carrier densities
\cite{JPS,TMF}.

To summarize, the crossover with carrier density and coupling changing
in this effective model with retarded interaction 
is smooth and the gap is not small
when the pair frequency is smaller than the boson frequency.
This approach can be used for studying the crossover
for the cases of different symmetries of the order parameter.

It is important to mention that the case considered here can be generalized 
on the case when one takes into account the vertex correction to the
electron-phonon interaction. This correction is usually small
when $e_F>>\omega_{0}$ (the Migdal theorem). 
It was shown in \cite{Ikeda},
that this correction is also small when $\epsilon_{F}<<\omega_{0}$.
However, in some cases the vertex correction can lead to strong
enhancement of superconductivity \cite{Pietronero1,Grimaldi1,Paci1}.
It is also necessary to note, that this correction even lead to enhancement 
of the $d$-wave superconductivity in the strongly correlated electron system
\cite{Paci1,Mierzejewski1},
despite phonon interaction due to the symmetry does not allow $d$-wave
pairing in the case when the short range electron repulsion is not considered.

\section{Fluctuations}

The role of fluctuations of the order parameter
in the 2D and even in the 3D case at $T=0$
can be significant. As it is shown in \cite{Kos1}, the Gaussian
fluctuation correction to the $s$-wave order parameter
is non-negligible even in the weak-coupling case. 
It was shown that the fluctuations
of the order parameter phase lead to increasing of the gap.

In this Section we show that 
the simultaneous order parameter modulus and phase
fluctuations in the model with local attraction
lead
to strong increasing of the order parameter 
at small carrier densities and to small
decreasing of the order parameter when the carrier densities are large.

At zero temperature the thermodynamic potential of the system
with local attraction has the following form:
$$
\Omega =v \left[\frac{|\Phi|^{2}}{V}-
\int\frac{d^{2}k}{(2\pi )^{2}}(\sqrt{\xi^{2}({\bf k})+|\Phi|^{2}}
-\xi ({\bf k}))\right].
$$ 
In other words, it depends on the combination
$|\Phi|^{2}=(\Re \Phi)^{2}+(\Im \Phi)^{2}$.
For studying fluctuations of the OP, 
it is convenient to use new real variables:
$$
{\bar \Phi} (x)=
\left(
\begin{array}{c}
\phi_{1}(x)\\
\phi_{2}(x)
\end{array}\right),
$$
such that 
${\bar \Phi}^{2}(x)=\phi_{1}^{2}(x)+\phi_{2}^{2}(x)
=|\Phi (x)|^{2}
$.
Another possibility is to choose the decomposition of the
order parameter 
on its phase and modulus $\Phi (x)=\Delta (x) \exp (i\theta (x))$.
However, it leads to some difficulties, since one needs
to keep the order parameter modulus 
positive at functional integration over the fluctuations.
The ``old'' order parameter 
variables are connected with the new ones in the following way:
$$
\Phi (x)=\phi_{1}(x)+i\phi_{2}(x), \ \ \ \ \ \ \ 
\Phi^{*} (x)=\phi_{1}(x)-i\phi_{2}(x).
$$
We assume that the mean-field value of the field ${\bar \Phi}$ 
is chosen as 
$$
{\bar \Phi}_{0} (x)=
\left(
\begin{array}{c}
\Delta\\
0
\end{array}\right).
$$

The order parameter ${\bar \Phi}$ can be written as
\begin{equation}
{\bar \Phi} (x)={\bar \Phi}_{0}+\delta{\bar \Phi} (x)\equiv
\left(
\begin{array}{c}
\Delta+\delta\phi_{1}(x)\\
\delta\phi_{2}(x)
\end{array}\right) 
\label{Phibar}
\end{equation}
in the case when its fluctuations are considered.
We neglect the fluctuations of the carrier density
$n_{f}({\bf r})$, and consider the homogeneous constant
value of $n_{f}$ over the lattice $n_{f}({\bf r})=n_{f}=$ const. 

Substitution of (\ref{Phibar}) into the expression
for the thermodynamic potential gives the following correction
to the thermodynamic potential in the second order in the fluctuations
\cite{Negele}:
\begin{eqnarray}
\delta\Omega =
-\int_{-\infty}^{\infty}\frac{d\nu}{(2\pi)}
\int\frac{d^2q}{(2\pi)^{2}}
\delta{\bar \Phi} (i\nu ,{\bf q})
{\hat A}(i\nu , {\bf q})
\delta{\bar \Phi} (-i\nu ,-{\bf q}),
\label{Gauss}
\end{eqnarray}
where ${\hat A}(i\nu , {\bf q})=\frac{1}{V}+\chi (i\nu , {\bf q})$ 
is a $2\times 2$ matrix
with the susceptibility components
$$
\chi_{jk}(i\nu , {\bf q})=\frac{1}{2}tr
\int_{-\infty}^{\infty}\frac{d\omega}{(2\pi)}
\int\frac{d^2k}{(2\pi)^{2}}
(-1)^{j+k}{\cal G}(i\omega_{+},{\bf k}_{+}){\hat \tau}_{j}
{\cal G}(i\omega_{-},{\bf k}_{+}){\hat \tau_{k}},
$$
where
$ 
\omega_{\pm}=\omega\pm \nu /2,
{\bf k}_{\pm}={\bf p}\pm {\bf q}/2$.
Integration over $\omega$ can be easily performed: 
$$
\chi_{11}(i\nu ,{\bf q})=
-\int\frac{d^2 k}{(2\pi )^{2}}
\frac{1}{2}\frac{E_{+}+E_{2}}{\nu^{2}+(E_{+}+E_{-})^{2}}
\left[ 
1+\frac{\xi_{+}\xi_{-}-\Delta^{2}}{E_{+}E_{-}}
\right],
$$
$$
\chi_{12}(i\nu ,{\bf q})=-\chi_{21}(i\nu ,{\bf q})
=
-\int\frac{d^2 k}{(2\pi )^{2}}
\frac{1}{2}\frac{E_{+}\xi_{-}+E_{-}\xi_{+}}{[\nu^{2}+(E_{+}+E_{-})^{2}]
E_{+}E_{-}},
$$
$$
\chi_{22}(i\nu ,{\bf q})=
-\int\frac{d^2 k}{(2\pi )^{2}}
\frac{1}{2}\frac{E_{+}+E_{2}}{\nu^{2}+(E_{+}+E_{-})^{2}}
\left[
1+\frac{\xi_{+}\xi_{-}+\Delta^{2}}{E_{+}E_{-}}
\right],
$$
where 
$E_{\pm}=\sqrt{\xi_{\pm}^{2}+\Delta^{2}}$,
$\xi_{\pm}=({\bf k}\pm {\bf q}/2)^{2}/(2m)-\mu$ is the free
fermion dispersion relation.
After the integrating out of the fluctuation field ${\bar \Phi}$
the correction to thermodynamic potential has the next form
$$
\delta\Omega =\frac{1}{2}
\int_{-\infty}^{\infty}\frac{d\nu}{(2\pi)}
\int\frac{d^2q}{(2\pi)^{2}}
ln \left[\{\frac{1}{V}+\chi_{11}(i\nu ,{\bf q})\}
    \{\frac{1}{V}+\chi_{22}(i\nu ,{\bf q})\}
   -\chi_{12}(i\nu ,{\bf q})\chi_{21}(i\nu ,{\bf q}) 
   \right].
$$
The factor $\sim \Delta$ which appears in the measure
of functional integration over $\Delta$ in $Z$ 
due to taking into account the symmetry
of the thermodynamic potential with respect to the
transformation ${\bar \Phi}\rightarrow e^{i\alpha}{\bar \Phi}$
(see, for example 
\cite{Negele}) is omitted in the last expression.
This factor can be absorbed in the measure of
the functional integral over $\Delta$ in the partition function, 
where the functional integration
can be actually performed over the variable $\Delta^{2}$.
Let us note, that only the first component 
$(\frac{1}{V}+\chi_{11}(i\nu ,{\bf q}))$ under the logarithm in $\delta\Omega$
will be present when
one considers the particular case of the order parameter
phase fluctuations.

It is useful to diagonalize the matrix ${\hat A}$, to find the contributions
which come from the phase and the modulus fluctuations of the order parameter.
Obviously, the first component will correspond to the phase fluctuations
and the second - for the modulus fluctuations, as it follows from the
definitions of the field ${\bar \Phi}$. 
So, it is easy to arrive to the following representation
$$
\delta\Omega =\frac{1}{2}
\int_{-\infty}^{\infty}\frac{d\nu}{(2\pi)}
\int\frac{d^2q}{(2\pi)^{2}}
ln \left[\{\frac{1}{V}+\chi_{\theta}(i\nu ,{\bf q})\}
    \{\frac{1}{V}+\chi_{\Delta}(i\nu ,{\bf q})\} 
   \right],
$$
where
$$
\chi_{\theta}(i\nu ,{\bf q})=
\frac{1}{2}\left[\chi_{11}(i\nu ,{\bf q})+\chi_{22}(i\nu ,{\bf q})\right]
-\sqrt{\frac{1}{4}\left[\chi_{11}(i\nu ,{\bf q})-\chi_{22}(i\nu ,{\bf q})
                  \right]^{2}
-\chi_{12}(i\nu ,{\bf q})\chi_{21}(i\nu ,{\bf q})
}
$$
$$
\chi_{\Delta}(i\nu ,{\bf q})=
\frac{1}{2}\left[\chi_{11}(i\nu ,{\bf q})+\chi_{22}(i\nu ,{\bf q})\right]
+\sqrt{\frac{1}{4}\left[\chi_{11}(i\nu ,{\bf q})-\chi_{22}(i\nu ,{\bf q})
                  \right]^{2}
-\chi_{12}(i\nu ,{\bf q})\chi_{21}(i\nu ,{\bf q})
}
$$
are the effective contributions to the thermodynamic potential
from the fluctuations
of the order parameter phase and the modulus, correspondingly.
 
The equations for the gap and for the chemical potential
(\ref{gap}) and (\ref{mu}) have the next form in the case
of the Gaussian correction to the thermodynamic 
potential due to the order parameter fluctuations:
$$
\frac{\Delta}{V}=\int\frac{d^2q}{(2\pi )^{2}}
\frac{\Delta}{2\sqrt{\xi ({\bf q})^{2}+\Delta^{2}}}
+\frac{1}{2}
\int_{-\infty}^{\infty}\frac{d\nu}{(2\pi)}
\int\frac{d^2q}{(2\pi)^{2}}
\left[
\frac{\partial \chi_{\theta}(i\nu ,{\bf q})/\partial\Delta}
{1/V+\chi_{\theta}(i\nu ,{\bf q})}
+\frac{\partial \chi_{\Delta}(i\nu ,{\bf q})/\partial\Delta}
{1/V+\chi_{\Delta}(i\nu ,{\bf q})}
\right],
$$
$$
n_{f}=\int\frac{d^2q}{(2\pi )^{2}}
\left[1-\frac{\xi ({\bf q})}{\sqrt{\xi ({\bf q})^{2}+\Delta^{2}}}
\right]
+\int_{-\infty}^{\infty}\frac{d\nu}{(2\pi)}
\int\frac{d^2q}{(2\pi)^{2}}
\left[
\frac{\partial \chi_{\theta}(i\nu ,{\bf q})/\partial\mu}
{1/V+\chi_{\theta}(i\nu ,{\bf q})}
+\frac{\partial \chi_{\Delta}(i\nu ,{\bf q})/\partial\mu}
{1/V+\chi_{\Delta}(i\nu ,{\bf q})}
\right].
$$

Considering the limit of weak coupling, the functions 
$\chi_{\theta}$ and $\chi_{\Delta}$
in the denominators can be neglected, than the integration
over $\nu$ can be performed and after some trivial manipulations
the following simple equation can be obtained:
\begin{eqnarray}
\frac{1}{V} =\int\frac{d {\bf q}}{(2\pi)^{2}}
\frac{1}{2\sqrt{\xi^{2}({\bf q})+\Delta^{2}}}
+\frac{V}{2}
\int\frac{d^2 k}{(2\pi )^{2}}
\int\frac{d^2 q}{(2\pi )^{2}}
\frac{\xi_{+}\xi_{-}}
{E_{+}^{3}E_{-}},
\label{gapequationfluc}
\end{eqnarray}
\begin{eqnarray}
n_{f}=\int\frac{d {\bf q}}{(2\pi)^{2}}
\left[1-\frac{\xi ({\bf q})}{\sqrt{\xi^{2}({\bf q})+\Delta^{2}}}\right]
+V\int\frac{d^2 k}{(2\pi )^{2}}
\int\frac{d^2 q}{(2\pi )^{2}}
\frac{\Delta^{2}(\xi_{+}-\xi_{-}) }
{E_{+}^{3}E_{-}}
\label{numberequationfluc}
\end{eqnarray}
(compare with (\ref{gaplocalT0}) and (\ref{mulocalT0})).
The substitution ${\bf q}\rightarrow {-\bf q}$
in a part of the terms was made in obtaining equations
(\ref{gapequationfluc}) and (\ref{numberequationfluc}).
It is interesting to note that in the case of the phase 
fluctuations the numerator under integral
in the last term of the the gap equation
will be $\frac{1}{2}\xi_{+}(\xi_{-}-\xi_{+})$ and
the last term in the number equation will be multiplied by
$\frac{1}{2}$.
It is also a good approximation to put $\mu =\epsilon_{F}$, since 
in the weak coupling regime
$\mu$
is different from $\epsilon_{F}$
only at very low carried densities.
The solution of the equation (\ref{gapequationfluc}) for the gap parameter
as function of $\epsilon_{F}$
at $\mu =\epsilon_{F}$ and 
different values of the coupling constant
is presented in Fig. 6.
\begin{figure}[h]
\centering{
\includegraphics[width=5.5cm,angle=270]{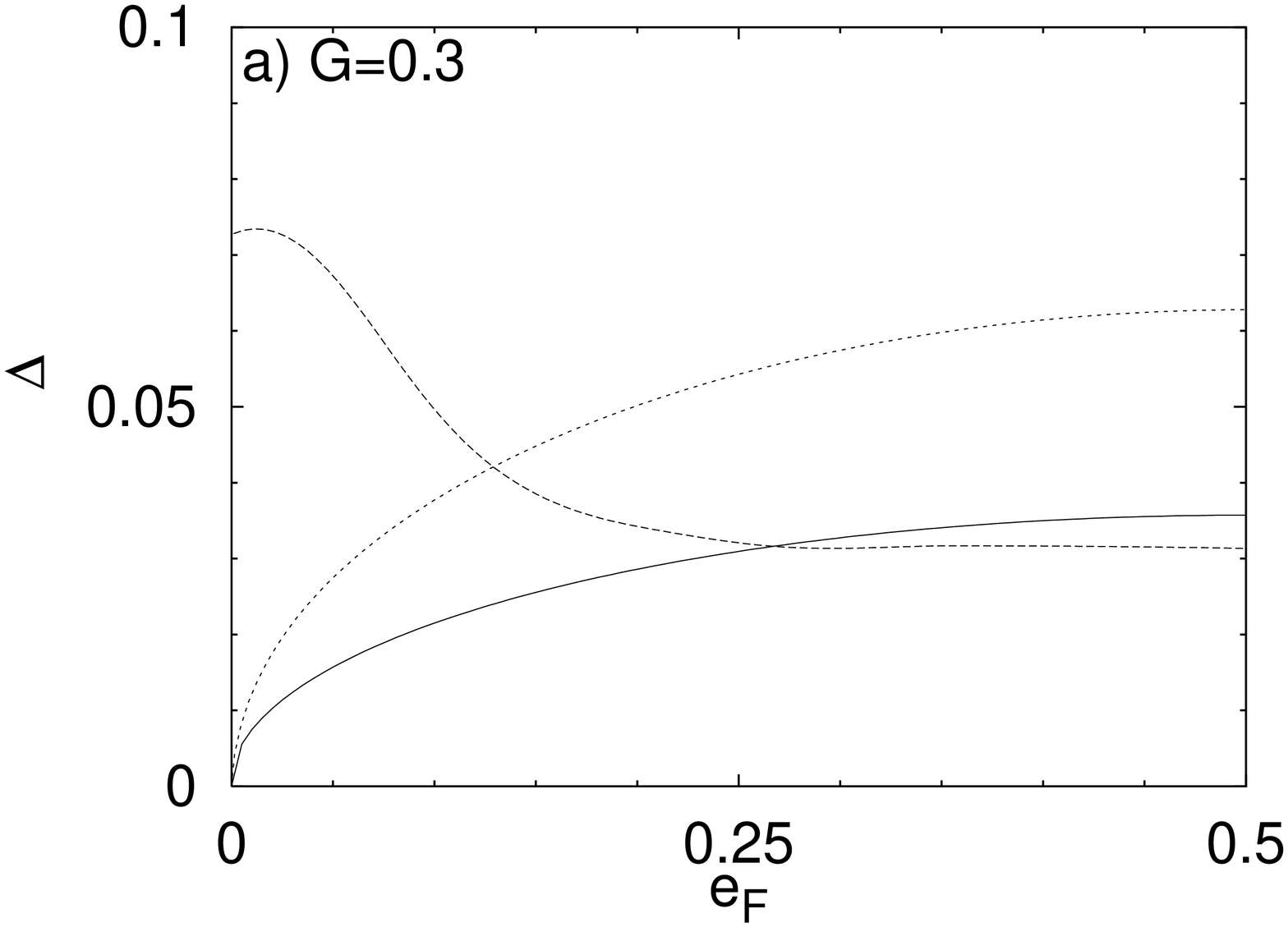}
\includegraphics[width=5.5cm,angle=270]{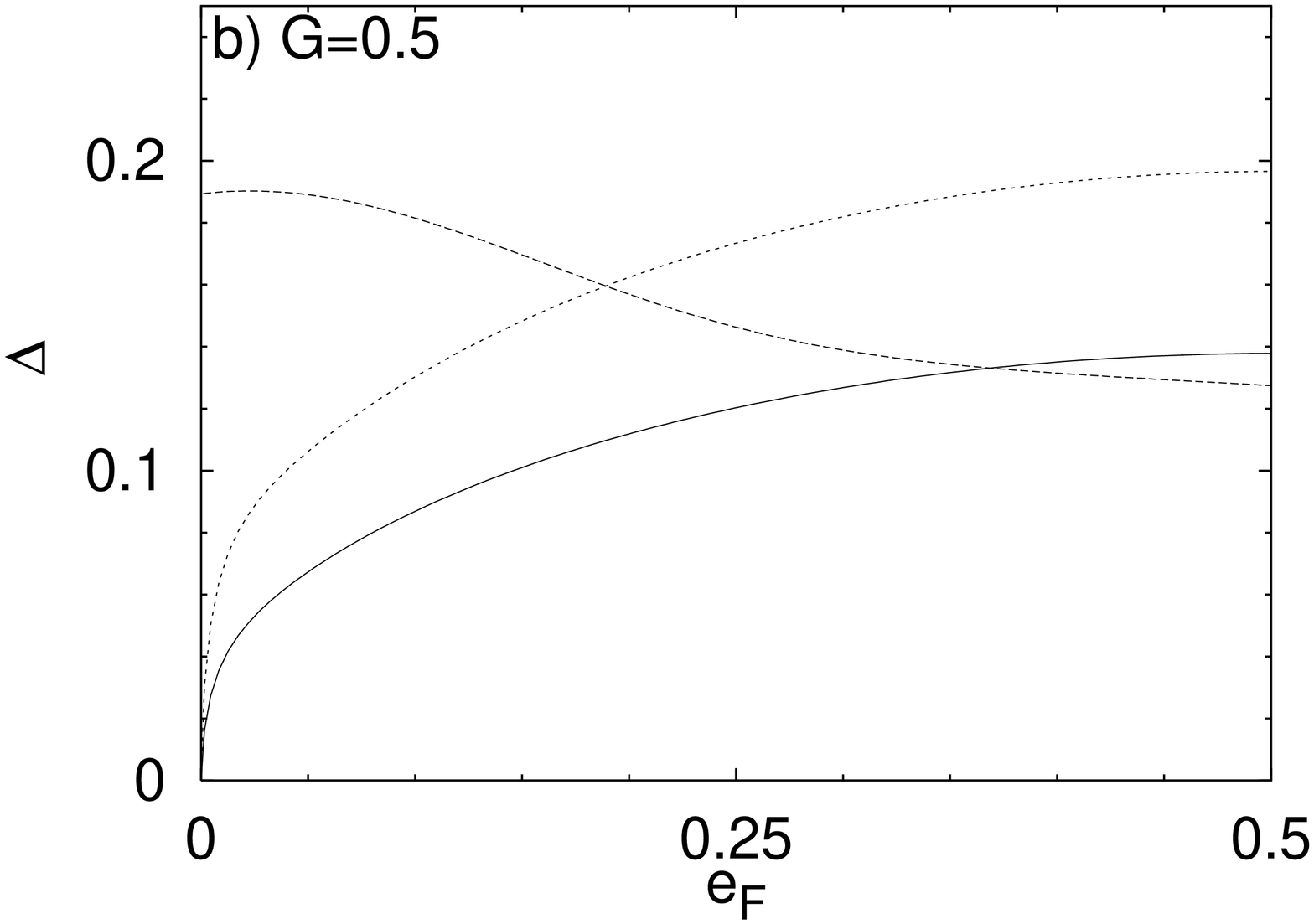}}
\caption{The dependence of $\Delta$ 
on $\epsilon_F$ 
is presented for the mean-field solution case (solid line) and for 
the order parameter 
fluctuations case (dashed line) at different values of the dimensionless
coupling parameter
$G=mV/(2\pi )$.
The dotted line is the estimation  of Ko$\check{s}$ 
and Millis \cite{Kos1} for the
case of the order parameter phase fluctuations.}  
\label{fig:6}
\end{figure}
The estimation of the order parameter in the case
of phase fluctuation is also presented. As it was shown in \cite{Kos1}
in the 2D case the phase fluctuations lead to an effective increasing
of the coupling constant $V\rightarrow V(1+\frac{2}{\pi^{2}})$.
The gap can be calculated from the standard mean-field BCS equation.
The comparison of different cases shows that the phase fluctuations
lead to the gap increasing, while the total fluctuations lead
to much stronger increasing of the gap at small carrier densities
and to decreasing of the gap when the carrier density is large.
The last result is familiar, but the first one is very surprising. 
The dependence of the gap on coupling at a small value of the
carrier density is presented in Fig.7.
These results suggest that the higher order corrections in fluctuations
should be studied to better understand the behavior of the system.

\begin{figure}[h]
\centering{\includegraphics[width=8.0cm,angle=270]{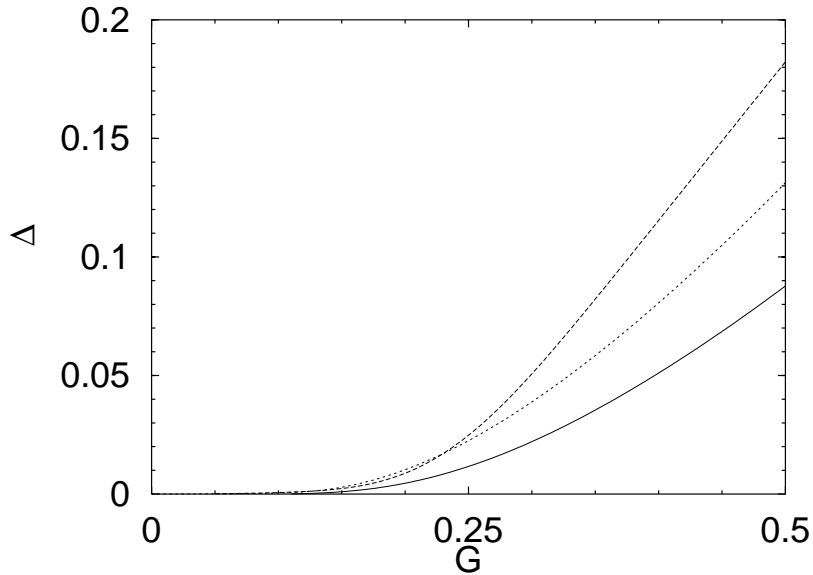}}
\caption{The dependence of $\Delta$ 
on $G$ is presented for mean field solution case (solid line) and for 
order parameter fluctuations case (dashed line) at $\epsilon_F=0.1$.
The dotted line is the estimation from \cite{Kos1} for the
order parameter phase fluctuations.}  
\label{fig:7}
\end{figure}

It should be mentioned that the role of the disorder due dopants
in the fluctuations 
of the inhomogeneous order parameter was studied by Yu.G.~Pogorelov and by 
the authors in \cite{Loktev1,Loktev2,IntJMP}.
We don't describe this important problem here, since it deserves
a special review.

\section{Conclusions}

In this paper the crossover from superfluidity to superconductivity
with doping increasing at $T=0$ in the cases of the
$s$-wave and $d$-wave pairing was reviewed.
In the 3D case this crossover does not take place
at small couplings, and the same situation takes place 
in the $d$-wave pairing case in two-dimensions, when the
interaction does not depend on the doping.
In the case when the correlation radius depends on doping
the minimal value of coupling for the two-particle
bound state exists even in the $s$-wave channel.
Also the gap can decrease with the doping in this case.

It was also shown, that the fluctuations of the order parameter
play an important role at $T=0$.
The fluctuations of the order parameter
phase in weak coupling limit in the case of the
$s$-wave pairing regime lead to increasing of superconductivity at
any physical carrier density, while the modulus fluctuations
lead to much stronger increasing of superconductivity at small
carrier densities, while at large carrier densities
they lead to suppression of the order parameter, 
as a result the gap is decreasing
in BCS regime, when both the modulus and the phase fluctuations
are taken into account.
This means that
the higher order fluctuation corrections must be investigated
in order to understand the self-consistent theory at low carrier
densities.

We would like to mention some direction which
can be interesting for the future investigations.
Studying of the problem of the crossover with realistic
dispersion relations is not performed even on the mean-field
level in many interesting cases.
The interplay between disorder and superconductivity,
and strong correlation and superconductivity is another
interesting topic for the investigations.

The fluctuations in the $d$-wave pairing channel and in other 
non-isotropic pairing channels even in the case
of Gaussian fluctuations were not studied carefully so far. 
It is also important to go beyond the Gaussian fluctuations,
since the pair susceptibility is diverging in the 2D and in the
3D cases, as it was mentioned in Ref.\cite{Kos1}. 
The role of the inter-layer coupling is another problem which
is not solved in general at the moment.
The solution of the problem mentioned above will lead
to better understanding of the superconducting properties
of systems with arbitrary carrier density and pairing potential.

\section*{Acknowledgments}

V.M.L. acknowledges partial support by SCOPES-project 7UKPJ062150.00/1 
of the Swiss National Science Foundation.

\end{document}